\documentclass[a4paper,oneside,10pt]{article}%
\usepackage{amsmath}
\usepackage{amsfonts}
\usepackage{amssymb}
\usepackage{graphicx}
\usepackage[square,numbers,sort&compress]{natbib}%
\providecommand{\U}[1]{\protect\rule{.1in}{.1in}}

\newtheorem{theorem}{Theorem}[section]

\newtheorem{definition}[theorem]{Definition}
\newtheorem{assumption}[theorem]{Assumption}
\newtheorem{example}[theorem]{Example}

\newtheorem{remark}[theorem]{Remark}

\numberwithin{equation}{section}

\begin{document}

\title{Explicit solutions for continuous time mean-variance portfolio selection with
nonlinear wealth equations}
\author{Shaolin Ji \thanks{Institute for Financial Studies, Shandong University, Jinan
250100, China and Institute of Mathematics, Shandong University, Jinan 250100,
China, Email: jsl@sdu.edu.cn. This work was supported by National Natural
Science Foundation of China (No. 11571203); Supported by the Programme of
Introducing Talents of Discipline to Universities of China (No.B12023). }
\and Xiaomin Shi\thanks{Corresponding author. Institute for Financial Studies,
Shandong University, Jinan 250100,
China. Email: shixm@mail.sdu.edu.cn, shixiaominhi@163.com. This
work was supported by National Natural Science Foundation of China (No.
11401091).}}
\maketitle

\textbf{Abstract}. This paper concerns the continuous time mean-variance
portfolio selection problem with a special nonlinear wealth equation. This
nonlinear wealth equation has a nonsmooth coefficient and the dual method
developed in \cite{Ji} does not work. We invoke the HJB equation of this
problem and give an explicit viscosity solution of the HJB equation.
Furthermore, via this explicit viscosity solution, we obtain explicitly the
efficient portfolio strategy and efficient frontier for this problem. Finally,
we show that our nonlinear wealth equation can cover three important cases.



\addcontentsline{toc}{section}{\hspace*{1.8em}Abstract}

\section{Introduction}

A mean-variance portfolio selection problem is to find the optimal portfolio
strategy which minimizes the variance of its terminal wealth while its
expected terminal wealth equals a prescribed level. Markowitz \cite{Ma1},
\cite{Ma2} first studied this problem in the single-period setting. It's
multi-period and continuous time counterparts have been studied extensively in
the literature; see, e.g. \cite{BJPZ}, \cite{JYZ}, \cite{LN}, \cite{LZL},
\cite{ZL} and the references therein. Most of the literature on mean-variance
portfolio selection focuses on an investor with linear wealth equation. But in
some cases, one need to consider nonlinear wealth equations. For example, a
large investor's portfolio selection may affect the return of the stock's
price which leads to a nonlinear wealth equation. When some taxes must be paid
on the gains made on the stocks, we also have to deal with a nonlinear wealth equation.

As for the continuous time mean-variance portfolio selection problem with
nonlinear wealth equation, Ji \cite{Ji} obtained a necessary condition for the
optimal terminal wealth when the coefficient of the wealth equation is smooth.
\cite{FLL} studied the continuous time mean-variance portfolio selection
problem with higher borrowing rate in which the wealth equation is nonlinear
and the coefficient is not smooth. They employed the viscosity solution of the
HJB equation to characterize the optimal portfolio strategy.

In this paper, the continuous time mean-variance portfolio selection problem
with a special nonlinear wealth equation is studied. This nonlinear wealth
equation has a nonsmooth coefficient and can cover the following three
important models: the first model is proposed by Jouini and Kallal \cite{JK}
and El Karoui et al \cite{EPQ1} in which an investor has different expected
returns for long and short position of the stock (see Example \ref{exm-1});
the second one is given in section 4 of \cite{CC} for a large investor (see
Example \ref{exm-2}); the third one is introduced in \cite{EPQ1} to study the
wealth equation with taxes paid on the gains (see Example \ref{exm-3}). We
invoke the Hamilton-Jacobi-Bellman (HJB for short) equation of this problem
and give an explicit viscosity solution of the HJB equation. Furthermore, via
this explicit viscosity solution, we obtain explicitly the efficient portfolio
strategy and efficient frontier for this problem.

The paper is organized as follows. In section 2, we formulate the problem. Our
main results are given in section 3. In section 4, we show that our wealth
equation (\ref{wealth}) can cover three important cases.

\section{Formulation of the problem}

Let $W$ be a standard 1-dimensional Brownian motion defined on a filtered
complete probability space $(\Omega,\mathcal{F},\{\mathcal{F}_{t}\}_{t\geq
0},P)$, where $\{\mathcal{F}_{t}\}_{t\geq0}$ denotes the natural filtration
associated with the $1$-dimensional Brownian motion $W$ and augmented. We
denote by $M^{2}(0,T)$ the space of all $\mathcal{F}_{t}-$progressively
measurable $\mathbb{R}$ valued processes $x$ such that $E\int_{0}^{T}x_{t}%
^{2}dt<\infty$.

We consider a financial market consisting of a riskless asset (the money
market instrument or bond) whose price is $S_{{}}^{0}$ and one risky security
(the stock) whose price is $S_{{}}^{1}$. An investor can decide at time
$t\in\lbrack0,T]$ what amount $\pi_{t}$ of his wealth $X_{t}$ to invest in the
stock. Of course, his decisions can only depend on the current information
$\mathcal{F}_{t}$, that is, the portfolio $\pi$ is $\mathcal{F}_{t}$-adapted.

For given deterministic continuous functions $r_{t},\underline{\theta}%
_{t},\bar{\theta}_{t},\sigma_{t}$ on $[0,T]$, consider the following nonlinear
wealth equation:
\begin{equation}%
\begin{cases}
dX_{t}=(r_{t}X_{t}+\pi_{t}^{+}\sigma_{t}\underline{\theta}_{t}-\pi_{t}%
^{-}\sigma_{t}\bar{\theta}_{t})dt+\pi_{t}\sigma_{t}dW_{t},\\
X_{0}=x_{0},\;t\in\lbrack0,T]
\end{cases}
\label{wealth}%
\end{equation}
where the functions $x^{+}:=\left\{
\begin{array}
[c]{c}%
x,\text{ if }x\geq0;\\
0,\text{ if }x<0,
\end{array}
\right.  $ and $x^{-}:=\left\{
\begin{array}
[c]{c}%
-x,\text{ if }x\leq0;\\
0,\text{ if }x>0.
\end{array}
\right.  $

We assume:

\begin{assumption}
\label{deter} $\underline{\theta}_{t}\geq0,\ \bar{\theta}_{t}\geq0$, a.e. on
$[0,T]$, $\sigma_{t}\neq0$, a.e. on $[0,T]$.
\end{assumption}

\begin{remark}
When $\underline{\theta}_{t}=\bar{\theta}_{t}$, a.e. on $[0,T]$, the wealth
equation (\ref{wealth}) reduces to the classical linear wealth equation.
\end{remark}

For a given expectation level $K$, consider the following continuous time
mean-variance portfolio selection problem:
\begin{align}
\label{optm} &  \mathrm{Minimize}\ VarX_{T}=E(X_{T}-K)^{2},\nonumber\\
&  s.t.%
\begin{cases}
EX_{T}=K,\\
\pi\in M^{2}(0,T),\\
(X,\pi)\ \ \mathrm{satisfies\ \ Eq.}(\ref{wealth}).
\end{cases}
\end{align}

Throughout the paper, we assume that $K\geq x_{0}e^{\int_{0}^{T}r_{s}ds}$.

The optimal strategy $\pi^{\ast}$ is called an efficient strategy. Denote the
optimal terminal value by $X_{T}^{\ast}$. Then, $(VarX_{T}^{\ast},K)$ is
called an efficient point. The set of all efficient points $\{(VarX_{T}^{\ast
},K)\mid K\in\lbrack x_{0}e^{\int_{0}^{T}r_{s}ds},+\infty)\}$ is called the
efficient frontier.

\begin{definition}
A portfolio $\pi$ is said to be admissible if $\pi\in M^{2}(0,T)$ and $(X,
\pi)$ satisfies $\mathrm{Eq}.(\ref{wealth})$.
\end{definition}

Denote by $\mathcal{A}(x_{0};0,T)$ the set of portfolio $\pi$ admissible for
the initial investment $x_{0}$. For simplicity, we set $\mathcal{A}%
(x_{0}):=\mathcal{A}(x_{0};0,T)$.

\bigskip

\section{Main results}

To deal with the constraint $EX_{T}=K$, we introduce a Lagrange multiplier
$-2\lambda\in\mathbb{R}$ and get the following auxiliary optimal stochastic
control problem:
\begin{align}
&  \mathrm{Minimize}\ E(X_{T}-K)^{2}-2\lambda(EX_{T}-K)=E(X_{T}-d)^{2}%
-(d-K)^{2}:=\hat{J}(\pi,d),\nonumber\label{optmau}\\
&  s.t.%
\begin{cases}
\pi\in M^{2}(0,T),\\
(X,\pi)\ \ \mathrm{satisfies\ \ Eq.}(\ref{wealth}),
\end{cases}
\end{align}
where $d:=K+\lambda$.

\begin{remark}
The link between problem (\ref{optm}) and (\ref{optmau}) is provided by the
Lagrange duality theorem (see Luenberger \cite{Lu})
\[
\min_{\pi\in\mathcal{A}(x_{0}),EX_{T}=K}VarX_{T}=\max_{d\in\mathbb{R}}%
\min_{\pi\in\mathcal{A}(x_{0})}\hat{J}(\pi,d).
\]

\end{remark}

So the optimal problem (\ref{optm}) can be divided into two steps. The first
step is to solve
\begin{equation}
\mathrm{Minimize}\ E(X_{T}-d)^{2},\ s.t.\ \pi\in\mathcal{A}(x_{0}),
\end{equation}
for any fixed $d\in\mathbb{R}$. The second step is to find the Lagrange
multiple which attains
\[
\max_{d\in\mathbb{R}}\min_{\pi\in\mathcal{A}(x_{0})}\hat{J}(\pi,d).
\]

To solve the first step, we introduce the stochastic control problem
\begin{equation}
v(t,x;d):=\inf_{\pi\in\mathcal{A}(x;t,T)}E(X_{T}-d)^{2},(t,x)\in
\lbrack0,T]\times\mathbb{R}%
\end{equation}
on $[t,T]$, subject to
\begin{equation}%
\begin{cases}
dX_{s}=(r_{s}X_{s}+\pi_{s}^{+}\sigma_{s}\underline{\theta}_{s}-\pi_{s}%
^{-}\sigma_{s}\bar{\theta}_{s})ds+\pi_{s}\sigma_{s}dW_{s},\\
X_{t}=x.
\end{cases}
\end{equation}

The value function $v(t,x;d)$ is a viscosity solution of the following HJB
equation (refer to \cite{YZ}):
\begin{equation}%
\begin{cases}
\frac{\partial v}{\partial t}+\inf\limits_{\pi\in\mathbb{R}}\big[\frac
{\partial v}{\partial x}(r_{t}x+\pi^{+}\sigma_{t}\underline{\theta}_{t}%
-\pi^{-}\sigma_{t}\bar{\theta}_{t})+\frac{1}{2}\frac{\partial^{2}v}{\partial
x^{2}}\sigma_{t}^{2}\pi^{2}\big]=0,\\
v(T,x;d)=(x-d)^{2}.
\end{cases}
\label{HJB}%
\end{equation}

Note that the functions $x^{+}$ and $x^{-}$ are nonsmooth. Then (\ref{HJB})
does not have a smooth solution. In the following theorem, we construct the
viscosity solution of (\ref{HJB}).

\begin{theorem}
Under Assumption \ref{deter}, the viscosity solution of the above HJB equation
(\ref{HJB}) is given by
\begin{equation}
v(t,x;d)=%
\begin{cases}
e^{-\int_{t}^{T}\underline{\theta}_{s}^{2}ds}(xe^{\int_{t}^{T}r_{s}ds}%
-d)^{2},\ \mathrm{if}\ x\leq de^{-\int_{t}^{T}r_{s}ds};\\
e^{-\int_{t}^{T}\overline{\theta}_{s}^{2}ds}(xe^{\int_{t}^{T}r_{s}ds}%
-d)^{2},\ \mathrm{if}\ x>de^{-\int_{t}^{T}r_{s}ds},
\end{cases}
\label{viscosity}%
\end{equation}
and the associated optimal feedback control is given by
\begin{equation}
\pi^{\ast}(t,x)=%
\begin{cases}
-\frac{\underline{\theta}_{t}}{\sigma_{t}}(x-de^{-\int_{t}^{T}r_{s}%
ds}),\ \mathrm{if}\ x\leq de^{-\int_{t}^{T}r_{s}ds};\\
-\frac{\bar{\theta}_{t}}{\sigma_{t}}(x-de^{-\int_{t}^{T}r_{s}ds}%
),\ \mathrm{if}\ x>de^{-\int_{t}^{T}r_{s}ds}.
\end{cases}
\label{OS-1}%
\end{equation}

\end{theorem}

\noindent\textbf{Proof:} Notice that the terminal condition of (\ref{HJB})
$(x-d)^{2}$ is a convex function in $x$. We conjecture that $v(t,x;d)$ is
convex in $x$ on $[0,T]$.

Then,
\begin{align*}
&  \ \ \ \ \ \inf\limits_{\pi\in\mathbb{R}}\big[\frac{\partial v}{\partial
x}(r_{t}x+\pi^{+}\sigma_{t}\underline{\theta}_{t}-\pi^{-}\sigma_{t}\bar
{\theta}_{t})+\frac{1}{2}\frac{\partial^{2}v}{\partial x^{2}}\sigma_{t}^{2}%
\pi^{2}\big]\\
&  =%
\begin{cases}
-\frac{1}{2}\frac{(\frac{\partial v}{\partial x})^{2}}{\frac{\partial^{2}%
v}{\partial x^{2}}}\underline{\theta}_{t}^{2}+\frac{\partial v}{\partial
x}r_{t}x,\ \mathrm{if}\ \frac{\partial v}{\partial x}\leq0;\\
-\frac{1}{2}\frac{(\frac{\partial v}{\partial x})^{2}}{\frac{\partial^{2}%
v}{\partial x^{2}}}\overline{\theta}_{t}^{2}+\frac{\partial v}{\partial
x}r_{t}x,\ \mathrm{if}\ \frac{\partial v}{\partial x}>0.
\end{cases}
\end{align*}

And the infimum in the above formula is attained at
\begin{equation}
\pi^{\ast}(t,x)=%
\begin{cases}
-\frac{\underline{\theta}_{t}\frac{\partial v}{\partial x}}{\sigma_{t}%
\frac{\partial^{2}v}{\partial x^{2}}},\ \mathrm{if}\ \frac{\partial
v}{\partial x}\leq0;\\
-\frac{\bar{\theta}_{t}\frac{\partial v}{\partial x}}{\sigma_{t}\frac
{\partial^{2}v}{\partial x^{2}}},\ \mathrm{if}\ \frac{\partial v}{\partial
x}>0.
\end{cases}
\label{OS-2}%
\end{equation}

The HJB equation (\ref{HJB}) becomes
\begin{equation}%
\begin{cases}
-\frac{\partial v}{\partial t}+\big[\frac{1}{2}\frac{(\frac{\partial
v}{\partial x})^{2}}{\frac{\partial^{2}v}{\partial x^{2}}}\underline{\theta
}_{t}^{2}-\frac{\partial v}{\partial x}r_{t}x\big]I_{\{\frac{\partial
v}{\partial x}\leq0\}}+\big[\frac{1}{2}\frac{(\frac{\partial v}{\partial
x})^{2}}{\frac{\partial^{2}v}{\partial x^{2}}}\overline{\theta}_{t}^{2}%
-\frac{\partial v}{\partial x}r_{t}x\big]I_{\{\frac{\partial v}{\partial
x}>0\}}=0,\\
v(T,x;d)=(x-d)^{2}.
\end{cases}
\label{HJB2}%
\end{equation}

We divide $[0,T]\times\mathbb{R}$ into three disjoint regions
\begin{align*}
&  \Gamma_{1}:=\{(t,x)\in\lbrack0,T]\times\mathbb{R}|x<de^{-\int_{t}^{T}%
r_{s}ds}\};\\
&  \Gamma_{2}:=\{(t,x)\in\lbrack0,T]\times\mathbb{R}|x>de^{-\int_{t}^{T}%
r_{s}ds}\};\\
&  \Gamma_{3}:=\{(t,x)\in\lbrack0,T]\times\mathbb{R}|x=de^{-\int_{t}^{T}%
r_{s}ds}\}.
\end{align*}
It is easy to verify that $v(t,x;d)$ defined in (\ref{viscosity}) is $C^{1,2}$
and satisfies (\ref{HJB2}) on $\Gamma_{1}$ and $\Gamma_{2}$.

On $\Gamma_{3}$,
\[
v(t,x;d)=\frac{\partial v}{\partial t}(t,x;d)=\frac{\partial v}{\partial
x}(t,x;d)\equiv0,
\]
Unfortunately, $\frac{\partial^{2}v}{\partial x^{2}}$ does not exist on
$\Gamma_{3}$ since $e^{\int_{t}^{T}(r_{s}-\underline{\theta}_{s}^{2}%
)ds}\not \equiv e^{\int_{t}^{T}(r_{s}-\overline{\theta}_{s}^{2})ds}$.

For any $\phi\in C^{\infty}([0,T]\times\mathbb{R})$, such that $(t,x)\in
\Gamma_{3}$ is a minimum point of $\phi-v$, it's easy to verify that
\[
\frac{\partial\phi}{\partial t}(t,x)=\frac{\partial\phi}{\partial x}(t,x)=0
\]
and
\[
\frac{\partial^{2}\phi}{\partial x^{2}}(t,x)\geq\text{max}\{2e^{\int_{t}%
^{T}(r_{s}-\underline{\theta}_{s}^{2})ds},2e^{\int_{t}^{T}(r_{s}%
-\overline{\theta}_{s}^{2})ds}\},\ (t,x)\in\Gamma_{3}.
\]
Then for any $\phi\in C^{\infty}([0,T]\times\mathbb{R})$, such that
$(t,x)\in\Gamma_{3}$ is a minimum point of $\phi-v$, we have
\begin{align*}
&  \ \ \ \frac{\partial\phi}{\partial t}+\inf\limits_{\pi\in\mathbb{R}%
}\big[\frac{\partial\phi}{\partial x}(r_{t}x+\pi^{+}\sigma_{t}%
\underline{\theta}_{t}-\pi^{-}\sigma_{t}\bar{\theta}_{t})+\frac{1}{2}%
\frac{\partial^{2}\phi}{\partial x^{2}}\sigma_{t}^{2}\pi^{2}\big]\\
&  =\frac{1}{2}\inf\limits_{\pi\in\mathbb{R}}\big[\frac{\partial^{2}\phi
}{\partial x^{2}}\sigma_{t}^{2}\pi^{2}\big]\\
&  \geq\frac{1}{2}\inf\limits_{\pi\in\mathbb{R}}\big[\text{max}\{2e^{\int%
_{t}^{T}(r_{s}-\underline{\theta}_{s}^{2})ds},2e^{\int_{t}^{T}(r_{s}%
-\overline{\theta}_{s}^{2})ds}\}\sigma_{t}^{2}\pi^{2}\big]\\
&  =0.
\end{align*}
Therefore, $v$ is a viscosity subsolution of the HJB equation (\ref{HJB}).

Similarly, for any $\phi\in C^{\infty}([0,T]\times\mathbb{R})$, such that
$(t,x)\in\Gamma_{3}$ is a maximum point of $\phi-v$, we have
\[
\frac{\partial\phi}{\partial t}(t,x)=\frac{\partial\phi}{\partial x}(t,x)=0.
\]
In this case,
\begin{align*}
&  \ \ \ \frac{\partial\phi}{\partial t}+\inf\limits_{\pi\in\mathbb{R}%
}\big[\frac{\partial\phi}{\partial x}(r_{t}x+\pi^{+}\sigma_{t}%
\underline{\theta}_{t}-\pi^{-}\sigma_{t}\bar{\theta}_{t})+\frac{1}{2}%
\frac{\partial^{2}\phi}{\partial x^{2}}\sigma_{t}^{2}\pi^{2}\big]\\
&  =\frac{1}{2}\inf\limits_{\pi\in\mathbb{R}}\big[\frac{\partial^{2}\phi
}{\partial x^{2}}\sigma_{t}^{2}\pi^{2}\big]\\
&  \leq0.
\end{align*}
Therefore, $v$ is a viscosity supersolution of the HJB equation (\ref{HJB}).
Finally, the terminal condition $v(T,x;d)=(x-d)^{2}$ is satisfied. By the
definition of viscosity solution, we know that $v(t,x;d)$ defined in
(\ref{viscosity}) is a viscosity solution of the HJB equation (\ref{HJB}). By
(\ref{OS-2}), it is easy to see that (\ref{OS-1}) holds.

This completes the proof. $\ \ \ \ \ \Box$

Now we determine the Lagrange multiple $d^{\ast}$ which attains $\max\limits
_{d\in\mathbb{R}}\min\limits_{\pi\in\mathcal{A}(x_{0})}\hat{J}(\pi,d).$

From (\ref{optmau}),
\begin{align*}
&  \ \ \ \ \min\limits_{\pi\in\mathcal{A}(x_{0})}\hat{J}(\pi,d)\\
&  =v(0,x_{0};d)-(d-K)^{2}\\
&  =%
\begin{cases}
e^{-\int_{0}^{T}\underline{\theta}_{s}^{2}ds}(x_{0}e^{\int_{0}^{T}r_{s}%
ds}-d)^{2}-(d-K)^{2},\ \mathrm{if}\ x_{0}\leq de^{-\int_{0}^{T}r_{s}ds};\\
e^{-\int_{0}^{T}\overline{\theta}_{s}^{2}ds}(x_{0}e^{\int_{0}^{T}r_{s}%
ds}-d)^{2}-(d-K)^{2},\ \mathrm{if}\ x_{0}>de^{-\int_{0}^{T}r_{s}ds}.
\end{cases}
\end{align*}
Set $d^{\ast}=\frac{K-x_{0}e^{\int_{0}^{T}(r_{s}-\underline{\theta}_{s}%
^{2})ds}}{1-e^{-\int_{0}^{T}\underline{\theta}_{s}^{2}ds}}$. We have
\begin{align}
&  \ \ \ \ \max_{d\in\mathbb{R}}\min\limits_{\pi\in\mathcal{A}(x_{0})}\hat
{J}(\pi,d)\nonumber\\
&  =\max_{d\in\mathbb{R}}[v(0,x_{0};d)-(d-K)^{2}]\nonumber\\
&  =v(0,x_{0};d^{\ast})-(d^{\ast}-K)^{2}\nonumber\\
&  =\frac{1}{e^{\int_{0}^{T}\underline{\theta}_{s}^{2}ds}-1}(K-x_{0}%
e^{\int_{0}^{T}r_{s}ds})^{2},
\end{align}
Therefore, the Lagrange multiple $\lambda^{\ast}=d^{\ast}-K=\frac
{K-x_{0}e^{\int_{0}^{T}r_{s}ds}}{e^{\int_{0}^{T}\underline{\theta}_{s}^{2}%
ds}-1}\geq0$.

The above analysis boils down to the following theorem.

\begin{theorem}
The efficient strategy of the problem (\ref{optm}) can be written as a
function of time $t$ and wealth $X$:
\begin{align}
\label{efficientstrategy}\pi^{\ast}(t,X)=%
\begin{cases}
-\frac{\underline{\theta}_{t}}{\sigma_{t}}(X-d^{\ast}e^{-\int_{t}^{T}r_{s}%
ds}),\ \mathrm{if}\ X\leq d^{\ast}e^{-\int_{t}^{T}r_{s}ds};\\
-\frac{\bar{\theta}_{t}}{\sigma_{t}}(X-d^{\ast}e^{-\int_{t}^{T}r_{s}%
ds}),\ \mathrm{if}\ X>d^{\ast}e^{-\int_{t}^{T}r_{s}ds}.
\end{cases}
\end{align}
Moreover, the efficient frontier is
\[
VarX_{T}=\frac{1}{e^{\int_{0}^{T}\underline{\theta}_{s}^{2}ds}-1}%
(K-x_{0}e^{\int_{0}^{T}r_{s}ds})^{2}\equiv\frac{1}{e^{\int_{0}^{T}%
\underline{\theta}_{s}^{2}ds}-1}(EX_{T}-x_{0}e^{\int_{0}^{T}r_{s}ds})^{2}.
\]

\end{theorem}

\begin{remark}
The efficient strategy (\ref{efficientstrategy}) indicates that the investor
should long the stock if his current wealth is less than $d^{*}e^{-\int%
_{t}^{T}r_{s}ds}$, otherwise he should take the short position.
\end{remark}

\bigskip

\section{Three examples}

In this section, three examples are given to show the applications of our main
results. The wealth equations in these examples are described by equation
(\ref{wealth}).

\begin{example}
\label{exm-1}Jouini and Kallal \cite{JK} and El Karoui et al \cite{EPQ1}
proposed the following model.

Under some circumstance, one has different expected returns for long and short
position of the stock. In this case, the assets prices are given by
\[%
\begin{cases}
dS_{t}^{0}=S_{t}^{0}r_{t}dt,\ S_{0}^{0}=s_{0};\\
dS_{t}^{1}=S_{t}^{1}\Big[\big(\underline{b}_{t}I_{\{\pi_{t}\geq0\}}+\bar
{b}_{t}I_{\{\pi_{t}<0\}}\big)dt+\sigma_{t}dW_{t}\Big],\ S_{0}^{1}=s_{1}>0;
\end{cases}
\]

Then the wealth process $X\equiv X_{{}}^{x,\pi}$ of the self-financed large
investor who is endowed with initial wealth $x_{0}>0$ is governed by the
following stochastic differential equation,
\[%
\begin{cases}
dX_{t}=\pi_{t}\frac{dS_{t}^{1}}{S_{t}^{1}}+(X_{t}-\pi_{t})\frac{dS_{t}^{0}%
}{S_{t}^{0}}\\
\ \ \ \ \ \ =(r_{t}X_{t}+\pi_{t}^{+}\sigma_{t}\underline{\theta}_{t}-\pi
_{t}^{-}\sigma_{t}\bar{\theta}_{t})dt+\pi_{t}\sigma_{t}dW_{t};\\
X_{0}=x_{0},
\end{cases}
\]
where $\underline{\theta}_{t}:=\frac{\underline{b}_{t}-r_{t}}{\sigma_{t}%
},\ \bar{\theta}_{t}:=\frac{\bar{b}_{t}-r_{t}}{\sigma_{t}}$ are risk premia
for long and short positions.
\end{example}

\begin{example}
\label{exm-2}Cuoco and Cvitanic \cite{CC} gave the following large investor model.

The portfolio strategy of a large investor can infect the expected return of
the stock.
\[%
\begin{cases}
dS_{t}^{0}=S_{t}^{0}r_{t}dt,\ S_{0}^{0}=s_{0};\\
dS_{t}^{1}=S_{t}^{1}\Big[\big(b_{t}-\varepsilon sgn(\pi_{t})\big)dt+\sigma
_{t}dW_{t}\Big],\ S_{0}^{1}=s_{1}>0;
\end{cases}
\]
where $\varepsilon$ is a given small positive number, and
\begin{equation}
sgn(x)=%
\begin{cases}
\frac{|x|}{x},\ \ \text{if}\ \ x\neq0;\\
0,\ \ \ \ \text{otherwise}.
\end{cases}
\end{equation}
In this specific large investor model, longing the risky security depresses
its expected return while shorting it increases its expected return as
explained in \cite{CC}.

The wealth equation can be written
\begin{align}
\label{wealthlarge}%
\begin{cases}
dX_{t} =(r_{t}X_{t}+(b_{t}-r_{t})\pi_{t}-\varepsilon|\pi_{t}|)dt+\pi_{t}%
\sigma_{t}dW_{t}\\
\ \ \ \ \ \ =(r_{t}X_{t}+\pi^{+}_{t}\sigma_{t}\underline{\theta}_{t}-\pi
_{t}^{-}\sigma_{t}\bar{\theta}_{t})dt+\pi_{t}\sigma_{t}dW_{t};\\
X_{0}=x_{0},
\end{cases}
\end{align}
where $\underline{\theta}_{s}:=\frac{b_{s}-r_{s}-\varepsilon}{\sigma_{s}}$ and
$\overline\theta_{s}:=\frac{b_{s}-r_{s}+\varepsilon}{\sigma_{s}}, \ s\in[0,T]$.
\end{example}


\begin{example}
\label{exm-3}El Karoui et al \cite{EPQ1} studied the following wealth equation
with taxes.

We suppose the assets prices are given by
\begin{align*}%
\begin{cases}
dS^{0}_{t}=S^{0}_{t} r_{t}dt, \ S^{0}_{0}=s_{0};\\
dS^{1}_{t}=S^{1}_{t}(b_{t}dt+\sigma_{t}dW_{t}), \ S^{1}_{0}=s_{1}>0.
\end{cases}
\end{align*}
And there are some taxes which must be paid on the gains made on the stock. In
this case, the wealth equation satisfies
\begin{align}
\label{wealthlarge}%
\begin{cases}
dX_{t} =(r_{t}X_{t}+(b_{t}-r_{t})\pi_{t}-\alpha\pi^{+}(b_{t}-r_{t}))dt+\pi
_{t}\sigma_{t}dW_{t}\\
\ \ \ \ \ \ =((r_{t}X_{t}+\pi^{+}_{t}\sigma_{t}\underline{\theta}_{t}-\pi
_{t}^{-}\sigma_{t}\bar{\theta}_{t})dt+\pi_{t}\sigma_{t}dW_{t};\\
X_{0}=x_{0},
\end{cases}
\end{align}
where $\underline{\theta}_{t}:=\frac{(1-\alpha)(b_{t}-r_{t})}{\sigma_{t}}$ and
$\bar\theta_{t}:=\frac{b_{t}-r_{t}}{\sigma_{t}}$, $\alpha\in[0,1)$ is a constant.
\end{example}

\renewcommand{\refname}

\end{document}